\documentstyle[aps,preprint]{revtex}
\begin{document}
\draft
\title{Quantum transport in mesoscopic systems and the measurement problem}
\author{S. A. Gurvitz}
\address{Department of Particle Physics, Weizmann Institute of
         Science, Rehovot 76100, Israel}
\date{\today}
\maketitle
\begin{abstract}
We study noninvasive measurement of stationary currents in mesoscopic systems.
It is shown that the measurement process is fully described by 
the Schr\"odinger equation without any additional reduction postulate and 
without the introduction of an observer. Nevertheless the possibility of 
observing a particular state out of coherent superposition leads 
to collapse of the wave function, even though the measured system 
is not distorted by interaction with the detector. 
Experimental consequences are discussed. 
\end{abstract}

According to the principles of quantum mechanics, a system
in a linear superposition of several states undergoes
a collapse to one of the states after measurement. More
precisely, the density matrix of the measured system,  
$\rho (t)=\sum_{nm}|n\rangle\rho_{m,n}\langle m|$ collapses  
to the statistical mixture, 
$\rho (t)\to\sum_{m}|m\rangle\rho_{m,m}\langle m|\,$. This is the
von Neumann projection postulate\cite{neu}.
Since both the measuring device (the {\em detector}) and the 
measured system are described by the Schr\"odinger equation, the question 
arises of how such a non-unitary process takes place.
The problem becomes
even more acute when the measured system is a macroscopic one\cite{leg}. 
Then the distortion due to interaction with the detector can be made 
negligibly small --- noninvasive measurement --- so that 
the collapse mechanism appears even more mysterious\cite{val}.

A weak point of many studies of the measurement problem is the lack 
of a detailed quantum mechanical treatment of the entire 
system, that is, of the detector and the measured system together. 
The reason is that 
the detector is usually a macroscopic system, the quantum mechanical 
analysis of which is very complicated. Mesoscopic systems 
might thus be more useful for study of the measurement problem. 
In this paper we discuss the measurement of stationary processes --- dc 
currents --- in mesoscopic systems by using the recently derived 
quantum rate equations for quantum transport\cite{glp1,glp2,gp}¤. 
These equations applied to the entire system,
allow us to follow the measurement process in great detail. 

Let us consider quantum transport in small tunneling structures 
(quantum dots). These systems have attracted great attention due to the 
possibility of 
investigating single-electron effects in the electric current\cite{likh}.
Until now research has been mostly concentrated on single dots, but 
the rapid progress in microfabrication technology has made it possible the
extension to coupled dot systems with aligned levels\cite{vdr,wbm}. 
In contrast with a single dot, the electron wave function inside
a coupled dot structure is a superposition of electron 
states localized in each of the dots. The collapse of the wave function and its
influence on the resonant current can thus be studied in these systems 
with a detector showing single electron charging  
of a quantum dot. Such a detector can be realized as   
a separate measuring circuit near the measured system\cite{pep,molen}. 
 
A possible setup is shown schematically in Fig. 1. Two quantum dots, 
represented by quantum wells, are coupled to two separate reservoirs 
at zero temperature. The resonant levels $E_0$ and $E_1$ are below the 
corresponding Fermi levels. 
In the absence of electrostatic interaction between electrons, the dc 
resonant currents in the upper well (the detector) and the lower well 
(the measured system) are respectively 
\begin{equation}
I_D^{(0)}=e\frac{\Gamma_0}{2},~~~~~~~~~~~~I_S^{(0)}=e\frac{\Gamma_L\Gamma_R}
{\Gamma_L+\Gamma_R}
\label{a1}
\end{equation}
The situation is different in 
the presence of electron-electron interaction between the dots, 
$H_{int}=Un_0n_1$, 
where $n_{0,1}$ are the occupancies of the upper and the lower wells
and $U$ is the Coulomb repulsion energy.  
If $E_{0}+U>\tilde E_F^L$, an electron from the left 
reservoir cannot enter the upper dot when the lower dot is 
occupied [Fig.~1~(c,d)]. On the other hand, the displacement of the level 
$E_1$ is much less important, since it remains below the Fermi level,
$E_{1}+U< E_F^L$. The upper dot can thus be considered as a detector,  
registering electrons entering the lower dot\cite{molen}. For instance, by 
measuring the detector current, 
$I_D$, one can determine the current in the lower dot. 
The same setup can be used for measurement of the current in the coupled dot
system shown in Fig. 2. 

The quantum transport in the structures described above 
is fully determined by evolution of the density matrix for the entire system:
$i\dot\rho (t)=[{\cal H},\rho ]$, for ${\cal H}=H_D+H_S+H_{int}$, 
where $H_{D,S}$ are the tunneling Hamiltonians of the
detector and the measured system, respectively, and 
$H_{int}=\sum_i U_in_0n_i$ describes
their mutual Coulomb interaction. $n_0$ and $n_i$ are the occupancies
of the detector and the dot $i$ of the multi-dot system. Using this 
equation one can find the current in the detector (or in the measured system) 
as the time derivative of the total average charge $Q(t)$ accumulated
in the corresponding right reservoir (collector): $I(t)=\dot Q(t)$, 
where $Q(t)=e$Tr$[\rho^R(t)]$, and $\rho^R(t)$ is the 
density matrix of the right reservoir. 
It was shown\cite{glp1,gp} that $I(t)$ is directly related to the 
density matrix of the multi-dot system $\sigma (t)$,
obtained from the total density matrix $\rho (t)$ by tracing out 
the reservoir states. One finds that the current in the detector or in the 
measured system is given by 
\begin{equation}  
I(t)=e\sum_c\sigma_{cc}(t)\Gamma^{(c)}_R, 
\label{a4}
\end{equation}
where $\sigma_{cc}\equiv \langle c|\sigma |c\rangle$
and the sum is over states $|c\rangle$ in which the well 
adjacent to the corresponding collector is occupied. 
$\Gamma_R^{(c)}$ is the partial width of the state $|c\rangle$
due to tunneling to the collector. The density 
matrix $\sigma_{ij}$ obeys the following system of modified rate 
equations \cite{gp},
\begin{mathletters}
\label{a7}
\begin{eqnarray}
\dot\sigma_{aa} & = &
 i\sum_{b(\neq a)}\Omega_{ab}(\sigma_{ab}-\sigma_{ba})
-\sigma_{aa} \sum_{d(\neq a)}\Gamma_{a \rightarrow d} +
 \sum_{c(\neq a)} \sigma_{cc}\Gamma_{c \rightarrow a}\;,
 \label{a7a}\\
\dot\sigma_{ab} & = & i(E_b - E_a) \sigma_{ab} +
i\left (\sum_{b'(\neq b)}\sigma_{ab'}\Omega_{b'b}
-\sum_{a'(\neq a)}\Omega_{aa'}\sigma_{a'b}\right )
\nonumber\\  &  & -\frac{1}{2} \sigma_{ab}
\left ( \sum_{d (\neq a)}\Gamma_{a \rightarrow d}
+\sum_{d (\neq b)}\Gamma_{b \rightarrow d}\right )
 +\frac{1}{2}\sum_{a'b'\neq ab}\sigma_{a'b'}
        \left (\Gamma_{a'\rightarrow a}+\Gamma_{b'\rightarrow b}\right ).
\label{a7b}
\end{eqnarray}
\end{mathletters}
These equations were
obtained from the many-body Schro\"odinger equation 
by integrating out the reservoir states, and assuming that the energy levels 
in the dots are not very close to the Fermi levels in the reservoirs.
Here $\Omega_{ij}$ denote hopping matrix elements 
of the tunneling Hamiltonian and $\sigma_{ba}=\sigma^*_{ab}$. 
The width $\Gamma_{a\to b}$ is the probability per unit time 
for the multi-dot system to make a transition from the state $|a\rangle$ to
the state $|b\rangle$ due to the tunneling to (or from) the
reservoirs, to interaction with the phonon bath, or to any other 
interaction generated by a continuum-state medium. 
The non-diagonal matrix elements are determined by Eq.~(\ref{a7b}), which 
resembles the optical Bloch equation supplemented with the last 
term. The latter contributes only for those 
{\em one-electron} transitions that convert at the same time 
the state $|a'\rangle$ into $|a\rangle$ {\em and} the state 
$|b'\rangle$ into $|b\rangle$.

It follows from Eq.~(\ref{a4}) that the current is determined
by the diagonal elements of the density matrix $\sigma (t)$.
The non-diagonal elements (``coherences'') 
influence the current via their coupling with the diagonal matrix
elements (the first term in Eq.~(\ref{a7a})). This coupling is due to   
transitions among discrete states. In the absence of such
transitions, for instance in the system shown in Fig. 1, 
the diagonal and non-diagonal
matrix elements are decoupled in the rate equations.
In this case the electron current is described by the classical rate
equations. Note, this does {\it not} imply that the coherences vanish.

Let us apply Eqs.~(\ref{a7}) to the system shown in Fig. 1.
For simplicity we disregard the spin degree of freedom.    
There are four available states of the device: 
$|a\rangle$ -- both wells are empty;
$|b\rangle$ -- the upper well is occupied; 
$|c\rangle$ -- the lower well is occupied;
$|d\rangle$ -- both wells are occupied. The rate equations 
for the diagonal matrix elements, obtained from Eq.~(\ref{a7a}), are:
\begin{mathletters}
\label{a2}
\begin{eqnarray}
\dot\sigma_{aa} & = &-(\Gamma_0+\Gamma_L)\sigma_{aa}
+\Gamma_0\sigma_{bb}+\Gamma_R\sigma_{cc}
\label{a2a}\\
\dot\sigma_{bb} & = &-(\Gamma_0+\Gamma'_L)\sigma_{bb}+\Gamma_0\sigma_{aa}
+\Gamma'_R\sigma_{dd}
\label{a2b}\\
\dot\sigma_{cc} & = &-\Gamma_R\sigma_{cc}+\Gamma_L\sigma_{aa}
+2\Gamma'_0\sigma_{dd}
\label{a2c}\\
\dot\sigma_{dd} & = &-(2\Gamma'_0+\Gamma'_R)\sigma_{dd}
+\Gamma'_L\sigma_{bb},
\label{a2d}
\end{eqnarray}
\end{mathletters}
which are supplemented with the initial condition,
$\sigma_{aa}(0)=1$ and $\sigma_{bb}(0)=\sigma_{cc}(0)=\sigma_{dd}(0)=0$.

The currents in the detector and in the lower well, Eq.~(\ref{a4}), are   
respectively $I_D(t)/e=\Gamma_0\sigma_{bb}(t)+\Gamma'_0\sigma_{dd}(t)$ and 
$I_S(t)/e=\Gamma_R\sigma_{cc}(t)+\Gamma'_R\sigma_{dd}(t)$. 
The stationary (dc) current corresponds to $I=I(t\to\infty )$. 

Solving Eqs. (\ref{a2}) we find in the limit $\Gamma_0,\Gamma'_0\gg
\Gamma_{L,R},\Gamma'_{L,R}$ 
\begin{equation}
I_S/e=\frac{\Gamma_R(\Gamma_L+\Gamma'_L)}{\Gamma_L+\Gamma'_L+2\Gamma_R}\;,
~~~~~~~~~~~~
\frac{I_D}{I_S}=\frac{\Gamma_0}{\Gamma_L+\Gamma'_L}\;,
\label{a5}
\end{equation}
The measurement of the detector current $I_D$ can thus be considered as 
a measurement of the current in the lower well, $I_S$. However,  
$I_S$, Eq. (\ref{a5}), is distorted by the detector
[$I_S \not =I_S^{(0)}$, Eq. (\ref{a1})] since $\Gamma'_L\not =\Gamma_L$.    
Nevertheless, the distortion can be made negligibly small. Indeed,
using the quasi-classical Gamow formula, $\Gamma_L=(1/T_1)\exp (-2S)$,
where $S=[2m(V-E_1)]^{1/2}L$ and $T_1=[2m/E_1]^{1/2}L_1$, one finds 
\begin{equation}
\frac{\Gamma'_L-\Gamma_L}{\Gamma_L}=\frac{U}{2E_1}+S\frac{U}{V-E_1}.
\label{a6}
\end{equation}
It follows that in the limit
$U/E_1\to 0$ and $U/(V-E_1)\to 0$, the width $\Gamma'_L\to\Gamma_L$. 
As a result $I_S\to I_S^{(0)}$, so that 
the measurement of $I_D$ can be considered as a {\em noninvasive}
measurement of the current $I_S$. 

In general, the measurement is a noninvasive one if 
the density matrix of the measured object does not depend on the 
detector parameters. One can easily see that this is precisely the case 
for the measurement discussed above. Indeed, by introducing
the density matrix of the measured system, $\bar\sigma (t)$,
obtained by tracing out the detector states in $\sigma (t)$,  
we find from Eq.~(\ref{a2}) in the limit $\Gamma'_{L,R}\to\Gamma_{L,R}$
\begin{mathletters}
\label{aa2}
\begin{eqnarray}
\dot{\bar\sigma}_{aa} & = &-\Gamma_L\bar\sigma_{aa}+\Gamma_R\bar\sigma_{cc}
\label{aa2a}\\
\dot{\bar\sigma}_{cc} & = &\Gamma_L\bar\sigma_{aa}-\Gamma_R\bar\sigma_{cc}
\label{aa2b}
\end{eqnarray}
\end{mathletters}
where $\bar\sigma_{aa}=\sigma_{aa}+\sigma_{bb}$ and  
$\bar\sigma_{cc}=\sigma_{cc}+\sigma_{dd}$.  
The detector parameters $\Gamma_0$ and $\Gamma'_0$  
are thus canceled in the equation of motion of the measured system.
This shows that a noninvasive measurement 
can be performed even if the measured system  
is not a macroscopic object\cite{leg,val}.

Note that the attachment of the detector to the measured system can be 
considered the last step in the measurement process. Even though one still 
need to register the current in the detector's right reservoir, this 
additional process can be carried out without any further distortion of 
the measured system and even the detector. Indeed, the reservoirs are 
systems with continuum spectrum, described by 
the classical rate equations that, as we have seen, admit a noninvasive 
measurement. 
 
Consider now resonant transport in the coupled well (coupled dot)
structure shown in Fig. 2. 
For simplicity, we assume that the Coulomb repulsion 
inside the double-dot is very strong, so 
only one electron can occupy it\cite{naz}. First consider the case of 
$E_0+U_1,\, E_0+U_2> \tilde E^L_F$. Then an electron cannot
enter the detector whenever another electron occupies the double-dot.  
All possible states of the measured system 
and the detector are shown in Fig. 2. We assume from the 
very beginning that the distortion of the
coupled-well parameters by the detector is negligibly small,
i.e., $\Gamma'_{L,R}\to \Gamma_{L,R}$, $\Omega'\to \Omega$. 
Using Eqs. (\ref{a7}) we can write the corresponding rate equations for 
the density matrix  
\begin{mathletters}
\label{a8}
\begin{eqnarray}
\dot\sigma_{aa} & = & -(\Gamma_L+\Gamma_0)\sigma_{aa}+\Gamma_0\sigma_{bb}+
\Gamma_R\sigma_{dd}
\label{a8a}\\
\dot\sigma_{bb} & = & -(\Gamma_L+\Gamma_0)\sigma_{bb}+\Gamma_0\sigma_{aa}+
\Gamma_R\sigma_{ff}
\label{a8b}\\
\dot\sigma_{cc} & = & i\Omega (\sigma_{cd}-\sigma_{dc})+\Gamma_L\sigma_{aa}+
2\Gamma'_0\sigma_{ee}
\label{a8c}\\
\dot\sigma_{dd} & = & -i\Omega (\sigma_{cd}-\sigma_{dc})-\Gamma_R\sigma_{dd}
+2\Gamma^{\prime\prime}_0\sigma_{ff}
\label{a8d}\\
\dot\sigma_{ee} & = & i\Omega (\sigma_{ef}-\sigma_{fe})-2\Gamma'_0\sigma_{ee}
+\Gamma_L\sigma_{bb}
\label{a8e}\\
\dot\sigma_{ff} & = & -i\Omega (\sigma_{ef}-\sigma_{fe})-(\Gamma_R+
2\Gamma^{\prime\prime}_0)\sigma_{ff}
\label{a8f}\\
\dot\sigma_{cd} & = & i\epsilon\sigma_{cd}+i\Omega (\sigma_{cc}-\sigma_{dd})
-\frac{1}{2}\Gamma_R\sigma_{cd}+(\Gamma'_0+
\Gamma^{\prime\prime}_0)\sigma_{ef}
\label{a8g}\\
\dot\sigma_{ef} & = & i(\epsilon+\Delta U)\sigma_{ef}+
i\Omega (\sigma_{ee}-\sigma_{ff})-(\Gamma'_0+\Gamma^{\prime\prime}_0+
\Gamma_R/2)\sigma_{ef},
\label{a8h}
\end{eqnarray}
\end{mathletters}
where $\epsilon =E_2-E_1$ and $\Delta U=U_2-U_1$. 
The currents in the detector ($I_D$)
and in the double-dot system ($I_S$) are given by Eq.~(\ref{a4}):
$I_D/e=\Gamma_0\sigma_{bb}+\Gamma'_0\sigma_{ee}+
\Gamma^{\prime\prime}_0\sigma_{ff}$, and  
$I_S/e=\Gamma_R(\sigma_{dd}+\sigma_{ff})$.
Solving Eqs.~(\ref{a8}) we find
\begin{equation}
I_S=I_S^{(0)}(1-\alpha I_S^{(0)}/\Gamma_0),~~~~~~~~~~~~~~~~~
I_D/I_S = \Gamma_0/2\Gamma_L, 
\label{a12}
\end{equation}  
where 
\begin{equation}  
I_S^{(0)}/e = \frac{\Gamma_R\Omega^2}{\displaystyle\epsilon^2+\Gamma_R^2/4+
\Omega^2(2+\Gamma_R/\Gamma_L)}
\label{a11}
\end{equation}
is the undistorted resonant current in the 
double-dot system for a strong Coulomb repulsion\cite{vdr,naz}. Consider 
$\Gamma_0\sim\Gamma'_0\sim{\Gamma^{\prime\prime}_0}\gg\Gamma_{L,R}, \Omega, 
\epsilon$. In this limit the coefficient $\alpha$ in Eq.~(\ref{a12}) is 
$\alpha=\Delta U(\Delta U+\epsilon)/(4\Gamma_0)^2$. As a result, 
$I_S\to I_S^{(0)}$ for $\Delta U\ll\Gamma_0$. Thus, the measurement 
of the detector current can be considered as a noninvasive measurement  
of the double-dot current. Also, one easily checks that
the density matrix of the
double-dot system, $\bar\sigma (t)$, is decoupled from
the detector in the limit $\Gamma'_{L,R}\to\Gamma_{L,R}$, 
$\Omega'\to\Omega$ and $\Delta U\to 0$.  Indeed, one finds from 
Eq.~(\ref{a8})
\begin{mathletters}
\label{a9}
\begin{eqnarray}
\dot{\bar\sigma}_{aa} & = & -\Gamma_L\bar\sigma_{aa}+\Gamma_R\bar\sigma_{dd}
\label{a9a}\\
\dot{\bar\sigma}_{cc} & = & i\Omega (\bar\sigma_{cd}-\bar\sigma_{dc})+
\Gamma_L\bar\sigma_{aa}
\label{a9b}\\
\dot{\bar\sigma}_{dd} & = & -i\Omega (\bar\sigma_{cd}-\bar\sigma_{dc})-
\Gamma_R\bar\sigma_{dd}
\label{a9c}\\
\dot{\bar\sigma}_{cd} & = & i\epsilon\bar\sigma_{cd} 
-i\Omega (\bar\sigma_{cc}-\sigma{dd})-\frac{1}{2}\Gamma_R\bar\sigma_{cd},
\label{a9d}
\end{eqnarray}
\end{mathletters}
where
$\bar\sigma_{aa}=\sigma_{aa}+\sigma_{bb}$,  $\bar\sigma_{cc}=\sigma_{cc}+
\sigma_{ee}$, $\bar\sigma_{dd}=\sigma_{dd}+\sigma_{ff}$ and $\bar\sigma_{cd}=
\sigma_{cd}+\sigma_{ef}$.

The above example shows that the behavior of the measured system is not  
distorted by the measurement, even if the system is in a linear 
superposition of different states, which 
affect of the detector in a different way (compare the states 
$c$ and $d$, or $e$ and $f$ in Fig. 2). Notice that 
the detector remains blocked, whenever an electron occupies any of the wells 
of the measured system. As a result such a measurement cannot single out  
the well in which an electron is located. 

Let us now increase the Fermi energy $\tilde E_F^L$, 
so that $E_0+U_2< \tilde E_F^L< E_0+U_1$.
In this case an electron from the left reservoir can
enter the detector even when the second dot of the measured system is occupied 
(the states $(d)$ and $(f)$ in Fig.2). As a result, 
the rate equations for $\sigma_{dd},\sigma_{ff},\sigma_{cd},\sigma_{ef}$
are changed. One finds    
\begin{mathletters}
\label{a13}
\begin{eqnarray}
\dot\sigma_{dd} & = & -i\Omega (\sigma_{cd}-\sigma_{dc})-(\Gamma_R+
\Gamma^{\prime\prime}_0)\sigma_{dd}+\Gamma^{\prime\prime}_0\sigma_{ff}
\label{a13d}\\
\dot\sigma_{ff} & = & -i\Omega (\sigma_{ef}-\sigma_{fe})-(\Gamma_R+
\Gamma^{\prime\prime}_0)\sigma_{ff}+\Gamma^{\prime\prime}_0\sigma_{dd}
\label{a13f}\\
\dot\sigma_{cd} & = & i\epsilon\sigma_{cd}+i\Omega (\sigma_{cc}-\sigma_{dd})
-\frac{1}{2}(\Gamma_R+\Gamma^{\prime\prime}_0)\sigma_{cd}+
\frac{1}{2}(\Gamma'_0+\Gamma^{\prime\prime}_0)\sigma_{ef}
\label{a13g}\\
\dot\sigma_{ef} & = & i(\epsilon +\Delta U)\sigma_{ef}+
i\Omega (\sigma_{ee}-\sigma_{ff})-\frac{1}{2}
(\Gamma_R+\Gamma'_0+2\Gamma^{\prime\prime}_0)\sigma_{ef},
\label{a13h}
\end{eqnarray}
\end{mathletters}
instead of Eqs.~(\ref{a8d}),(\ref{a8f})-(\ref{a8h}). 
Such a modification does not affect
the diagonal density matrix of the measured system, 
$\bar\sigma_{ii}(t)$, Eqs.~(\ref{a9a})-(\ref{a9c}). 
However,   
the rate equation for the non-diagonal density matrix 
element, $\bar\sigma_{cd}$, is different from Eq.~(\ref{a9d}).  
We obtain from Eqs.~({\ref{a13g})-({\ref{a13h}) for $\Delta U=0$
\begin{equation}
\dot{\bar\sigma}_{cd}  =  i\epsilon\bar\sigma_{cd} 
-i\Omega (\bar\sigma_{cc}-\sigma_{dd})-\frac{1}{2}\Gamma_R\bar\sigma_{cd}
-\frac{1}{2}(\Gamma^{\prime\prime}_0\bar\sigma_{cd}+\Gamma'_0\bar\sigma_{ef})
\label{a14}
\end{equation}
Thus, the widths $\Gamma'_0$, $\Gamma^{\prime\prime}_0$ 
are not canceled out in 
the density matrix of the double-dot system, despite the fact that no 
parameter of this system is distorted by the detector.
Moreover, the influence of the detector on the measured
current, $I_S$, is very strong. Indeed, take for simplicity 
$\Gamma^{\prime\prime}_0\simeq\Gamma'_0$. Then $I_S=I_S^{(1)}$, where 
\begin{equation}  
I_S^{(1)}/e = \frac{\Gamma_R\Omega^2}{\displaystyle\epsilon^2\Gamma_R/(\Gamma_R
+\Gamma'_0)+\Gamma_R(\Gamma_R+\Gamma'_0)/4+\Omega^2(2+\Gamma_R/\Gamma_L)}
\label{a15}
\end{equation}
It follows from Eq.~(\ref{a15}) that $I_S^{(1)}\to 0$ for 
$\Gamma'_0\gg \Gamma_{L,R}$, $\Omega$, $\epsilon$. However, the current
in the double-dot system returns to a previous (non-distorted) 
value, $I_S^{(0)}$, Eq.~(\ref{a11}), for $\tilde E_F^L>E_0+U_1$, Fig. 3. 
(We predict the same behavior of the current $I_S$ as a function of
$\tilde E_F^L$ if the detector is located near the second dot).

The expected strong drop-off of the current $I_S$ for
$E_0+U_1>\tilde E_F^L>E_0+U_2$ can be interpreted as 
an ``observation'' effect, yet without an
``observer''. Indeed, this configuration of the Fermi level 
allows us, in principle, to distinguish a particular dot occupied by 
an electron, whereas other configurations do not. 
It is clear that this effect 
cannot be explained by an interaction between
electrons inside the detector and the measured system, since the distortion  
of the measured system due to this interaction is negligibly small. 
Even if the distortion of the system cannot be totally neglected, 
the same distortion would also exist for  
$\tilde E_F^L<E_0+U_1$ or $\tilde E_F^L>E_0+U_2$, where no effect is expected. 

Actually, our quantum rate equations point to the origin of the 
observation effect. Consider first Eq.~(\ref{a7a}), describing the diagonal 
density  matrix elements. One finds that for each transition ${a\to a'}$ 
there exists the reverse ${a'\to a}$, which contributes with the opposite 
sign to the rate equation for $\sigma_{a'a'}$. Therefore the corresponding 
rates $\Gamma$ cancel in the rate equation for $\sigma_{aa}+\sigma_{a'a'}$. 
As a result, the detector rates drop out from the density matrix
when the latter is traced over the detector states (provided the rates 
of the measured system are not distorted by the measurement).
However, the rates $\Gamma$ are not in balance
in the rate equation for non-diagonal density-matrix elements, Eq.~(\ref{a7b}). 
One easily finds that the negative contributions from the 
transitions $(ab)\to (a'b)$ and $(ab)\to (ab')$ have no positive counterparts 
(in contrast with the transitions $(ab)\to (a'b'))$\cite{gp}.
This is precisely the case of measurement of a system in a coherent 
superposition when one of the superposed states can generate or prevent 
transitions between continuum and isolated states in the detector 
while the other state does not. As a result, the corresponding 
detector rates are not canceled in the density matrix describing the measured 
system, even though the distortion of this system is negligibly small. 

Although we concentrated in this paper on the measurement of currents
in mesoscopic systems, we assert that the results are valid for  
the measurement problem in general. First, consider the detector and 
the role of an observer. Since we found no 
dividing line between a microscopic and a macroscopic (classical) description, 
the detector need not be a macroscopic object. 
We require only the absence of transitions between discrete states, so that 
the diagonal and nondiagonal density matrix elements 
of the detector are decoupled. In this case the subsequent (noninvasive) 
interactions with an observer do not change the detector behavior,
and therefore cannot influence the wave function collapse. 
It follows that the result of the measurement is not affected by the observer.

With respect to measurements of systems in a linear superposition, 
we found that the measurement process is fully reproduced by 
the Schr\"odinger equation, written as quantum rate equations for 
the density matrix, without an independent projection postulate\cite{neu}. 
Indeed, one easily obtains from Eqs.~(\ref{a9}),(\ref{a14}) that 
$\bar\sigma_{cd}\to 0$ for $\Gamma'_0\to\infty$, so that the density matrix 
of the double-dot system collapses into the corresponding statistical 
mixture. Notice, however, that for a finite $\Gamma'_0$, the nondiagonal 
density matrix elements $\bar\sigma_{cd}, \bar\sigma_{dc}$ do not disappear, 
but are only diminished. Therefore the damping of nondiagonal density matrix 
elements due to the environment does not necessarily 
lead to their elimination for 
$t\to\infty$, as suggested in\cite{zur}. 
  
Although the measurement process is described by the Schr\"odinger equation,
the observation paradox is still there. It appears that  
a very sensitive detector, far away 
from the measured system but still capable of observing different states 
from the linear superposition, would strongly affect the measured system. 
This situation resembles the EPR paradox, but some features 
are different. First, the above observation paradox appears as
a stationary state phenomenon. Second, we do not need any special 
initial correlations 
between the electrons in the detector and the measured system. The most 
important difference is the possibility of influencing directly the measured 
current by switching the detector on (or off). Such a process can 
also be studied using our rate equations (\ref{a7}), describing  
time dependence of the density matrix, but special 
attention should be paid to the relativistic treatment\cite{aharonov}. 
\section*{Acknowledgments}
I am grateful to Ya. Prager and B. Svetitsky for reading the manuscript 
and making valuable comments on it.

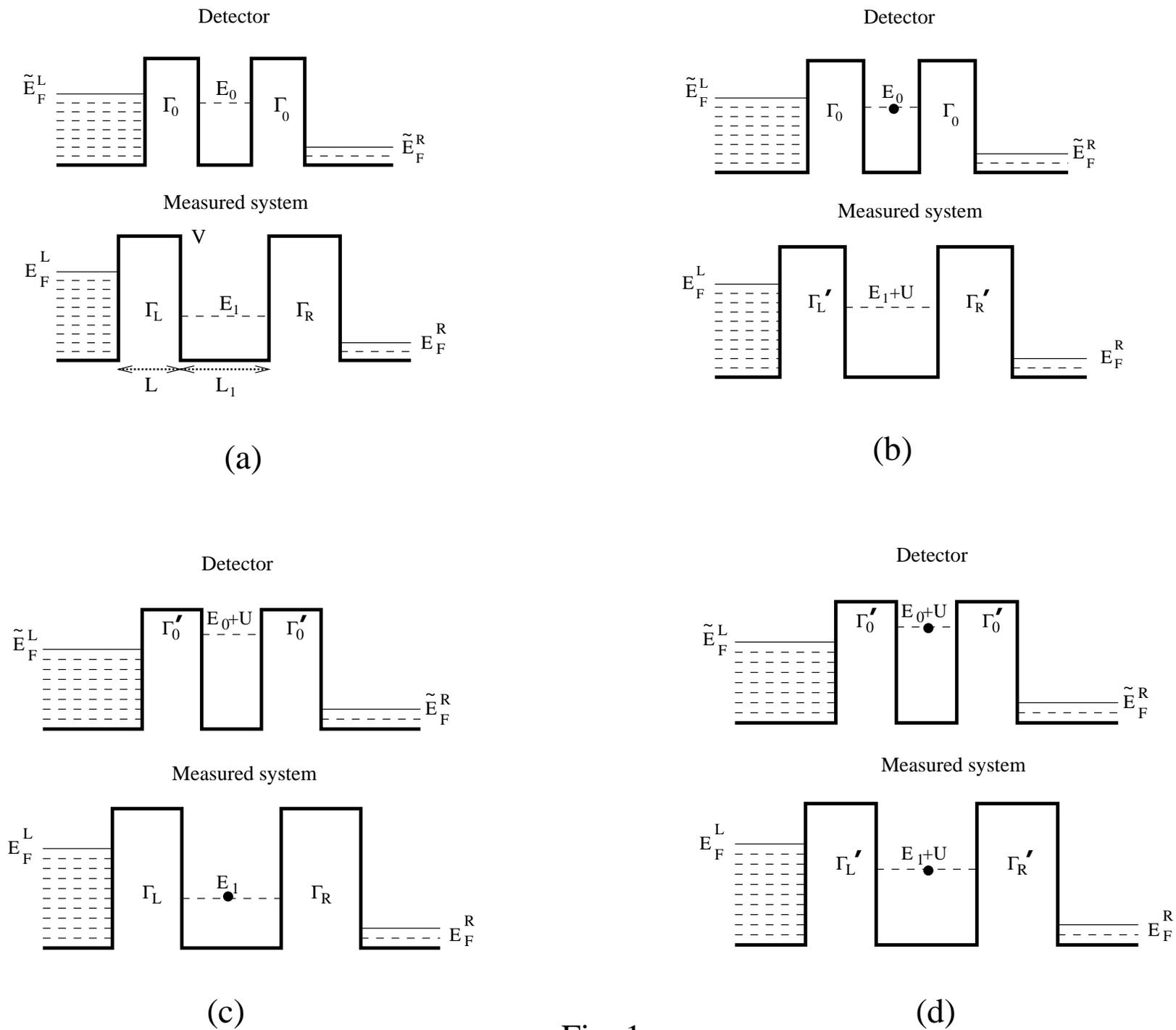
\begin{figure}[h]
\caption{The measurement of resonant current in a single-dot structure by
another, nearby dot. 
All possible electron states of the detector (the upper well)
and the measured system (the lower well) are shown. Also indicated are 
the tunneling rates ($\Gamma$), the left barrier height ($V$), 
barrier width ($L$), and width ($L_1$) of the lower well.}
\end{figure}
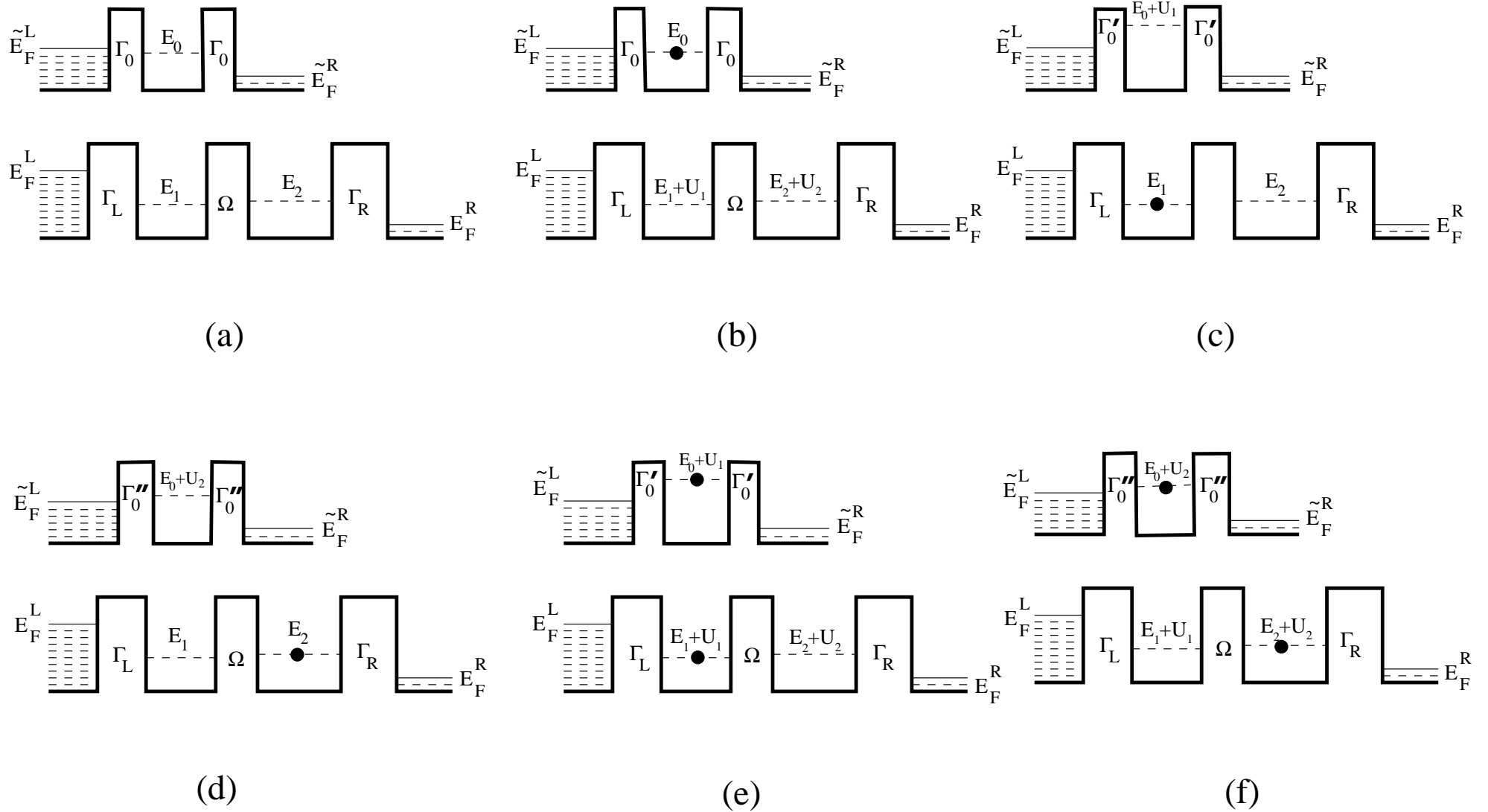
\begin{figure}[h]
\caption{The measurement of resonant current in a double-dot structure.
The energy level of the upper dot (the detector) is above the Fermi level 
whenever an electron occupies one of the dots in the
double-dot structure. $\Omega$ is the coupling between the dots.}
\end{figure}
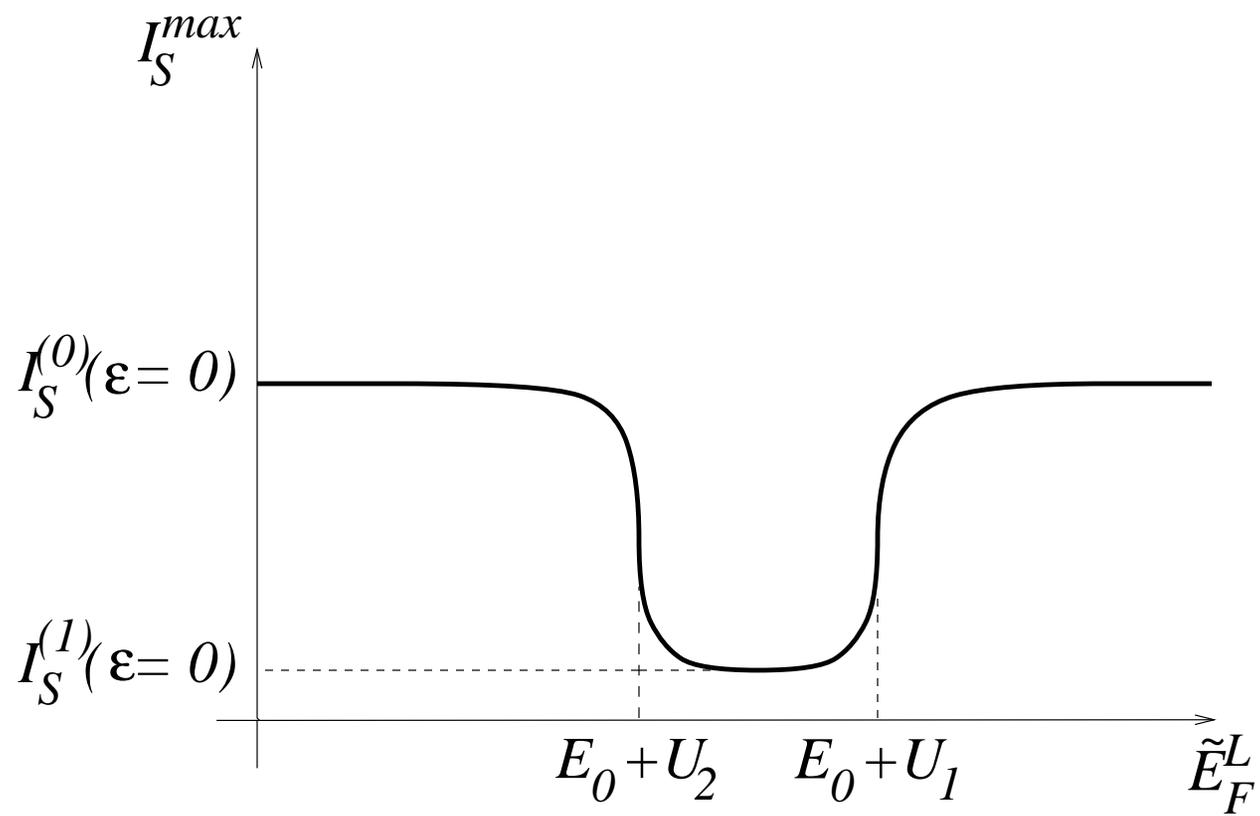
\begin{figure}[h]
\caption{Maximal current in the double dot structure
$I^{max}_S=I_S(\epsilon =0)$ as a function of the Fermi energy of the 
left reservoir adjacent to the detector. The detector does not distort any 
parameters of the measured system.}
\end{figure}


\begin{references}
\bibitem{neu} J. von Neumann, {\em Mathematische Grundlagen der Quantentheorie}
(Springer, Berlin, 1931). The statistical mixture means that the measured 
system is actually in one of its states, although there is no way to 
determine in which particular state the system is. Otherwise, quantum 
mechanics would be a pure deterministic theory.
\bibitem{leg} A.J. Leggett and A. Garg, Phys.\ Rev.\ Lett. 
{\bf 54}, 857 (1985); 
Phys.\ Rev.\ Lett. {\bf 59}, 1621 (1987);  
Phys.\ Rev.\ Lett. {\bf 63}, 2159 (1989). 
The predicted violation of the temporal Bell inequalities 
has not yet been observed. 
\bibitem{val} For a realization of interaction-free measurement see
L. Vaidman, Quant. Opt. {\bf 6}, 119 (1994); P. Kwiat, H. Wenfurter, 
T. Herzog, A. Zeilinger and M. Kasevich, 
Phys.\ Rev.\ Lett. {\bf 74}, 4763 (1995).
\bibitem{glp1} S.A. Gurvitz, H.J. Lipkin, and Ya.S. Prager, 
Mod.\ Phys.\ Lett.\ B {\bf 8}, 1377 (1994). 
\bibitem{glp2} S.A. Gurvitz, H.J. Lipkin, and Ya.S. Prager, 
Phys. Lett. A, in press. 
\bibitem{gp} S.A. Gurvitz and Ya.S. Prager, cond-mat/9511026, Phys. Rev. B,
in press. 
\bibitem{likh} D.V. Averin and K.K. Likharev, in 
{\em Mesoscopic Phenomena in Solids}, edited by B. Altshuler, 
P.A. Lee, and R.A. Webb (Elsevier, Amsterdam, 1991).
\bibitem{vdr} N.C. van der Vaart, S.F. Godijn, Y.V. Nazarov,
C.J.P.M. Hartmans, J.E. Mooij, L.W. Molenkamp, and C.T. Foxon, 
Phys.\ Rev.\ Lett. {\bf 74}, 4702 (1995).
\bibitem{wbm} F.R. Waugh, F.R. Waugh, M.J. Berry, D.J. Mar, R.M. Westervelt,
K.L. Campman and A.C. Gossard,
Phys.\ Rev.\ Lett. {\bf 75}, 4702 (1995).
\bibitem{pep} M. Field, C.G. Smith, M. Pepper, D.A. Ritchie, J.E.F. Frost, 
G.A.C. Jones, and D.G. Hasko,  
Phys. Rev. Lett. {\bf 70}, 1311 (1993).
\bibitem{molen} L.W. Molenkamp, K. Flensberg and M. Kemerink,  
Phys. Rev. Lett. {\bf 75}, 4282 (1995).
\bibitem{naz} Yu.V. Nazarov, Physica B {\bf 189}, 57 (1993),
T.H. Stoof and Yu.V. Nazarov, Phys. Rev. B {\bf 55}, 1050 (1996).
\bibitem{zur} W.H. Zurek, Physics Today {\bf 44}, No. 10, 36 (1991); 
{\em ibid}, {\bf 46}, No. 4, 13 (1993), and references therein.
\bibitem{aharonov} Y. Aharonov, D. Albert and L. Vaidman, 
Phys. Rev. D {\bf 34}, 1805 (1986).

\end{references}
\end{document}